# A New Approach for ARMA Pole Estimation Using Higher-Order Crossings

Timothy I. Salsbury, *Member, IEEE* and Ashish Singhal, *Member, IEEE*

***Abstract*—The paper describes a new method for estimating the poles of an ARMA model using higher-order crossings. The method involves transforming counts of crossing events into estimates of ARMA poles via the autocorrelation domain. An important advantage of the method is that the crossing counts are the only features that need to be stored from the original data. The poles of an ARMA model of a control loop correspond to the roots of the characteristic equation and are thus useful for evaluating control performance.**

## I. Introduction

Many processes including feedback control loops can be modeled as white noise passing through a dynamic system [1]. The process is ARMA (Auto-Regressive Moving Average) when the dynamic system is a linear rational discrete transfer function and the noise signal is an independent and identically distributed (IID) sequence of random variables.

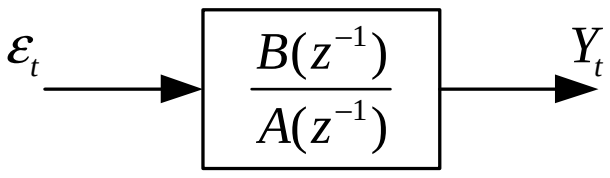

Figure 1: An Example SISO ARMA process

Figure 1 shows a general ARMA process where $\varepsilon_t$ is the random variable and:

$$\begin{aligned} B(z^{-1}) &= 1 + b_1 z^{-1} + b_2 z^{-2} + \ldots + b_m z^{-m} \\ A(z^{-1}) &= 1 + a_1 z^{-1} + a_2 z^{-2} + \ldots + a_n z^{-n} \end{aligned} \quad (1)$$

where $z^{-1}$ is used here as a backward-shift operator such that: $Y_t z^{-d} \equiv Y_{t-d}$.

Identification of an ARMA model includes two main parts: (1) finding an appropriate order; and (2) estimating parameters values. The focus of this paper is on the second part and it is assumed that the user specifies an appropriate order. Methods for ARMA parameter estimation are described in basic textbooks on time series analysis, e.g., [2]-[4]. A problem with the textbook methods is that they are not well suited for on-line application in low-cost devices. This is because the methods either require storage of a large batch of data and/or require some form of non-linear optimization [5].

 T. I. Salsbury and A. Singhal are with the Controls Research Group of Johnson Controls, Inc., 507 E Michigan Street, Milwaukee, WI 53202 USA. (phone: 414-524-4660; fax: 414-524-5810; e-mail: timothy.i.salsbury@jci.com).

Recently, new methods for performance assessment have been developed for control systems that require estimation of an ARMA model [6], [7]. A typical assumption is that these methods will be used in conjunction with an off-line time series analysis program that will process a batch of data. However, in low-cost industries such as building automation, access to these programs can be limited and it may not always be possible to store enough data. It is therefore difficult for these industries to take advantage of the new technologies. Recursive approaches to ARMA estimation are available, but they usually involve local linearization [8], [9] or high-order AR approximation [10]. The linearization approach can be susceptible to robustness problems and false minima [11], [12], while the AR approach will perform poorly if the order is not properly selected [13].

One way to address the non-linearity issue associated with ARMA parameter estimation is to split the problem into two parts: estimation of the numerator $B(z^{-1})$ and estimation of the denominator $A(z^{-1})$. By adopting an approach based on autocorrelation lags, $A(z^{-1})$ can be calculated by means of the modified (or extended) Yule Walker equations [14]-[16]. In control applications, the denominator is often of more interest than the numerator because it represents the characteristic equation of a feedback loop. Although the use of autocorrelation lags facilitates a linear problem formulation for denominator estimation, traditional approaches for estimating these lags still require storing and processing a batch of data.

This paper proposes using high-order crossings as a way to obtain autocorrelation lags without having to store a batch of data. A computationally efficient and low-storage procedure for ARMA pole estimation is introduced based on combining the higher-order crossings and modified Yule-Walker techniques. The use of zero-crossings represents a frequency-based approach to ARMA estimation and is therefore related to other frequency-based methods such as those described in [17]. The paper demonstrates how the method can be used to detect oscillatory modes in a SISO feedback control loop.

## II. Estimation of ARMA Poles using Autocorrelation Lags

For the ARMA process shown in Figure 1, the signal $Y_t$ is given by:

$$Y_t = \frac{B(z^{-1})}{A(z^{-1})}\varepsilon_t \qquad (2)$$

It is well known that the parameters in an ARMA($n$,$m$) process of the type in (2) can be related to a sequence of autocorrelation lags by taking expectations [3]. In particular, the parameters in the autoregressive part (i.e., denominator) of an ARMA model are related linearly to the sequence of autocorrelation lags as follows:

$$\begin{aligned} A(z^{-1})\rho_k &= 0 \\ \rho_k &= -a_1\rho_{k-1} - a_2\rho_{k-2} - \ldots - a_n\rho_{k-n} \end{aligned} \qquad (3)$$

for $k > m$. The parameters in the denominator of an ARMA($n,m$) model can therefore be determined from the autocorrelation lag-$k$ values by solving a set of simultaneous equations. This approach has been referred to as the *modified* or *extended* Yule-Walker method in [14]-[16], where it has been used to estimate AR models in the presence of colored noise.

Given sample estimates $\hat{\rho}_k, \ldots \hat{\rho}_{k+n}$ of the autocorrelation lag values for a signal $Y_t$, the coefficients in the denominator of the generating ARMA model can be estimated from:

$$\hat{A} = \hat{\Omega}^{-1}\hat{P}, \quad \text{for } k > m \qquad (4)$$

where:

$$\hat{\mathrm{P}} = \begin{bmatrix} \hat{\rho}_k \\ \hat{\rho}_{k+1} \\ \vdots \\ \hat{\rho}_{k+n-1} \end{bmatrix}; \quad \hat{A} = \begin{bmatrix} -\hat{a}_1 \\ -\hat{a}_2 \\ \vdots \\ -\hat{a}_n \end{bmatrix}; \text{ and}$$

$$\hat{\Omega} = \begin{bmatrix} \hat{\rho}_{k-1} & \hat{\rho}_{k-2} & \cdots & \hat{\rho}_{k-n} \\ \hat{\rho}_k & \hat{\rho}_{k-1} & \cdots & \hat{\rho}_{k-n+1} \\ \vdots & \vdots & \ddots & \vdots \\ \hat{\rho}_{k+n-2} & \hat{\rho}_{k+n-3} & \cdots & \hat{\rho}_{k-1} \end{bmatrix} \qquad (5)$$

The poles, $z = p_{z,i}$, of the ARMA model are obtained by solving for the roots of the equation $A(z) = z^n + \hat{a}_1 z^{n-1} + \ldots + \hat{a}_n$.

A traditional method for calculating the autocorrelation lags for a uniformly sampled signal is to collect a batch of data and calculate $\hat{\rho}_k$ at lag $k$ from:

$$\hat{\rho}_k = \frac{\sum_{t=1}^{q-|k|}(Y_{t-|k|} - \mu)(Y_t - \mu)}{\sum_{t=1}^{q}(Y_t - \mu)^2} \qquad (6)$$

where $-q < k < q$ and $\mu = \frac{1}{q}\left(\sum_{t=1}^{q} Y_t\right)$. Online calculation of $\hat{\rho}_k$ using this approach requires storing a batch of data. In this paper, we present an alternative method for calculating the autocorrelation lags that does not require storing a batch of data. The method is based on higher-order crossings and is explained in the following section.

## III. Review of zero-crossings theory

Kac [18] and Rice [19] were among the first to derive relationships between the frequency at which a stationary signal crosses its expected value and spectral properties. Rice's work, in particular, led to a mathematical framework for zero-crossings analysis that has since been formalized in standard textbooks, e.g., Cramér and Leadbetter [20]. More recently, Kedem [21], [22], [23] advanced the theory by providing a thorough treatment of discrete signals and by introducing the concept of higher-order crossings.

Zero-crossings analysis is mostly restricted to stochastic signals that are stationary and Gaussian. The assumption of Gaussianity leads to certain simplifications, but other distributions can be treated via minor modifications [24]. The basic idea behind zero-crossings analysis is simply to count the number of times a signal crosses its expected value. Because amplitude is not needed for the analysis, a clipped version of the signal can be considered instead. For example, if $Y_1, \ldots, Y_N$ is a zero-mean stationary time series, a clipped series $\{X_t\}$ is created from:

$$X_t \equiv \begin{cases} 1, & Y_t \geq 0 \\ 0, & Y_t < 0 \end{cases}, \quad t = 1, \ldots, N \qquad (7)$$

An expression for the number of zero-crossings in the original series $\{Y_t\}$ is then:

$$D_1 \equiv \sum_{t=2}^{N}(X_t - X_{t-1})^2 \qquad (8)$$

where $D_1$ is the number of zero-crossings. The subscript on $D$ is used here as an index to distinguish higher-order crossings, which are explained in the next section. An important theoretical result is the relation between the number of zero-crossings and the lag-one autocorrelation for a stationary Gaussian time series. This relation is referred to as the "cosine formula" and is given by:

$$\rho_1 = \cos\left(\frac{\pi E[D_1]}{N-1}\right) \qquad (9)$$

where an early derivation of (9) can be found in [25]. Given a finite sample of data, an estimate of the lag-one autocorrelation can be made by substituting the actual number of zero-crossings counted in the data for the expected value.

## IV. Higher-Order Crossings

Further signal properties can be obtained from zero-crossing counts on filtered versions of the original data series. In principle, any linear filter can be applied to construct new relationships between signal properties and zero crossing counts. In this paper, we use a differencing operation, which is a very simple high-pass filter. Repeated application of filter operations and the associated zero-crossings counts are referred to as "higher-order crossings", or HOC. Application of a difference operation is denoted by:

$$\nabla Y_t \triangleq Y_t - Y_{t-1} = (1 - z^{-1}) Y_t \tag{10}$$

where $\nabla$ is the difference operator. In general, the $k^{th}$ difference of $\{Y_t\}$ is:

$$\nabla^k Y_t = (1 - z^{-1})^k Y_t = \sum_{j=0}^{k} \binom{k}{j} (-1)^j Y_{t-j} \tag{11}$$

It is now useful to define $D_k$ as the number of zero crossings in the series $\nabla^{k-1} Y_t$. The HOCs have some interesting properties, such as their monotonicity where: $0 \le E[D_1] \le E[D_2] \le \ldots \le (N-1)$. Also, if $\omega$ is the highest frequency in the spectrum, $\pi E[D_j]/(N-1) \to \omega$ as $j \to \infty$. For any signal that is mixed with white noise, $\omega$ is then always the Nyquist frequency. In contrast, when only a single frequency exists (e.g., a pure sine wave), all $E[D]$ values are equal.

For a Gaussian signal, it can be proven that higher-order crossings uniquely determine the spectral distribution function up to a constant [23]. One representation of the relationship between the spectral distribution function and HOCs obtained from differencing operations is:

$$\cos\left(\frac{\pi E[D_{k+1}]}{N-1}\right) = \frac{\int_{-\pi}^{\pi} \cos(\omega)(\sin\omega/2)^{2k} \, dF(\omega)}{\int_{-\pi}^{\pi} (\sin\omega/2)^{2k} \, dF(\omega)} \tag{12}$$

where $F(\omega)$ is the spectral density function. Higher-order crossings can also be related to a signal's autocorrelation sequence and a derivation of this relationship can be found in [23]. The relationship is reproduced here without proof:

$$\cos\left(\frac{\pi E[D_{k+1}]}{N-1}\right) = \frac{-\binom{2k}{k-1} + \rho_1\left[\binom{2k}{k} + \binom{2k}{k-2}\right] - \cdots + (-1)^k \rho_{k+1}}{\binom{2k}{k} - 2\rho_1 \binom{2k}{k-1} + \cdots (-1)^k 2\rho_k} = \frac{\nabla^{2k}\rho_{k-1}}{\nabla^{2k}\rho_k} \tag{13}$$

Equation (13) thus enables calculation of a signal's autocorrelation sequence from an HOC series $D_1, \ldots, D_{k+1}$. Re-arrangement of (13) yields:

$$\rho_{k+1} = (-1)^k \left[\Psi \cos\left(\pi E[\tilde{D}_{k+1}]\right) - \Phi\right] \tag{14}$$

where $\tilde{D}_{1\ldots k-1} = D_{1\ldots k-1}/(N-1)$ are the set of higher-order crossings, normalized by the sample size; also:

$$\Psi = \binom{2k}{k} - 2\rho_1 \binom{2k}{k-1} + 2\rho_2 \binom{2k}{k-2} - \ldots + (-1)^k 2\rho_k$$
$$\Phi = -\binom{2k}{k-1} + \rho_1\left[\binom{2k}{k} + \binom{2k}{k-2}\right] - \rho_2\left[\binom{2k}{k-1} + \binom{2k}{k-3}\right] + \ldots + (-1)^{k-1}\rho_k \binom{2k}{1} \tag{15}$$

Equation (14) therefore enables lag-*k* autocorrelation values to be estimated from an HOC sequence.

## V. Summary of the Method

The proposed method is computationally simple and only requires storage of the number of times a considered signal and its differences cross their expected values. The method can be broken down into four main computational steps as described below.

1. Count the number of times an observed signal, e.g., $Y_t$, and its differences cross their expected values and create the normalized HOC sequence: $\{\tilde{D}_1, \ldots, \tilde{D}_{m+n}\}$.
2. Estimate the autocorrelation lags from the sequence of HOC using (14).
3. Estimate the parameters in the denominator of the ARMA process by solving the modified Yule-Walker equations given in (4).
4. Find the roots of the estimated ARMA denominator using a standard root finding algorithm.

The proposed approach is best illustrated by an example. For simplicity we consider an AR process where $B(z^{-1}) = 1$ and $A(z^{-1}) = 1 + a_1 z^{-1} + a_2 z^{-2}$. The only difference to the procedure when MA terms are included is that more lags need to be estimated. Let the output of the AR process be $Y_t$ and assume that crossing counts have been made on the output up to some point in time. The raw output signal is therefore transformed into a set of normalized crossing counts such that: $\{Y_{1\ldots T}\} \Rightarrow \{\tilde{D}_1, \tilde{D}_2\}$. Note that only two crossing counts are required because the process is ARMA(2,0). For an ARMA(*n*,*m*) process, the total number of counts would be $n + m$.

The normalized zero-crossing counts $\{\tilde{D}_1, \tilde{D}_2\}$ are the only variables that need to be calculated directly from the data. In an on-line set-up, this is useful because these counts require very little computation and storage. All other calculations could be carried out off-line, either periodically, or on demand. Once the zero crossing counts have been obtained, the next step is to calculate the autocorrelation lags using Equation (14). For the two required lags, the equations resolve to:

$$\hat{\rho}_1 = \cos(\pi \tilde{D}_1) \tag{16}$$

$$\begin{aligned}\hat{\rho}_2 &= (-1)^1 \left[ \left( \binom{2}{1} - 2\hat{\rho}_1 \binom{2}{0} \right) \cos(\pi \tilde{D}_2) - \left(-\binom{2}{0} + \hat{\rho}_1 \binom{2}{1}\right) \right] \\ &= (2\hat{\rho}_1 - 2)\cos(\pi \tilde{D}_2) - 1 + 2\hat{\rho}_1 \end{aligned} \quad (17)$$

Once the autocorrelation lags have been calculated, the modified Yule-Walker equations can be used to estimated the coefficients in $A(z^{-1})$ using Equation (4), i.e.,

$$\begin{pmatrix} -\hat{a}_1 \\ -\hat{a}_2 \end{pmatrix} = \begin{pmatrix} 1 & \hat{\rho}_1 \\ \hat{\rho}_1 & 1 \end{pmatrix}^{-1} \begin{pmatrix} \hat{\rho}_1 \\ \hat{\rho}_2 \end{pmatrix} \quad (18)$$

Estimates of the coefficients in $A(z^{-1})$ are thus obtained by solving a set of simultaneous equations. The roots of $A(z^{-1})$ correspond to the poles of the ARMA process and would be calculated for this example by solving the standard quadratic expression.

An important practical advantage of the zero-crossings approach is that it is resistant to the presence of extreme values in amplitude because it uses information only from the frequency domain.

The algorithm could also be easily modified to include exponential forgetting of past data by adjusting the way in which the crossing counts are updated. For example, the *D* values can be obtained by calculating an EWMA of the number of samples between crossing events. The reciprocal of this average period is then an estimate of the normalized zero-crossing frequency $\tilde{D}$. An EWMA estimate of the period between zero-crossings would be calculated from:

$$\bar{T}_j = (1-\lambda)\bar{T}_{j-1} + \lambda T_j \quad (19)$$

where $0 \le \lambda \le 1$ is a forgetting factor, $T_j$ is the number of samples between the previous and current crossing events, and $\bar{T}_j$ is EWMA of the period. An estimate of the average number of zero-crossings per sample is then simply: $\tilde{D} = 1/\bar{T}_j$.

## VI. Example Results

This section presents results of using the method to estimate the poles in three different ARMA processes, one being a SISO feedback control loop.

### A. ARMA(1,1) and ARMA(2,1) Estimation

In this section, the method was applied to signals generated by passing 10,000 samples of an IID Gaussian noise sequence through two example transfer functions:

$$\begin{aligned} G_{1,1}(z^{-1}) &= \frac{Y_t}{\varepsilon_t} = \frac{1-0.5z^{-1}}{1-0.95z^{-1}} \\ G_{2,1}(z^{-1}) &= \frac{Y_t}{\varepsilon_t} = \frac{1-0.5z^{-1}}{(1-0.9z^{-1})(1-0.95z^{-1})} \\ &= \frac{1-0.5z^{-1}}{1-1.85z^{-1}+0.855z^{-2}} \end{aligned} \quad (20)$$

where $G_{1,1}(z^{-1})$ and $G_{2,1}(z^{-1})$ yield ARMA(1,1) and ARMA(2,1) processes respectively.

Table 1: Results of ARMA(1,1) and ARMA(2,1) estimation

| **System** | $G_{1,1}(z^{-1})$ | $G_{2,1}(z^{-1})$ |
|---|---|---|
| *Number of Samples* | 10000 | 10000 |
| HOC sequence $\{D_1; D_2; D_3\}$) | 2088; 6338; __ | 268; 3249; 6355 |
| Lags from HOC $\{\hat{\rho}_1; \hat{\rho}_2; \hat{\rho}_3\}$ | 0.792; 0.754; __ | 0.996, 0.989, 0.979 |
| Lags from (6) $\{\hat{\rho}_1; \hat{\rho}_2; \hat{\rho}_3\}$ | 0.784; 0.742; __ | 0.996; 0.987; 0.976 |
| Poles from HOC $\{r_1; r_2\}$ | 0.952; __ | 0.968; 0.870 |
| Poles from (6) $\{r_1; r_2\}$ | 0.946; __ | 0.962; 0.870 |
| RMSE† (HOC) | 0.028 | 0.154 |
| RMSE† ( (6)) | 0.047 | 0.078 |

† RMSE $\triangleq \sqrt{\frac{1}{N\hat{\sigma}_Y^2}\sum_{t=1}^{N}\left(Y_t - \hat{Y}_t\right)^2}$, $\hat{Y}_t$ is the predicted value and $\hat{\sigma}_Y^2$ is the estimated sample variance of $Y_t$.

The method was applied to the signals $Y_t$ from the two processes in order to determine the poles in the transfer functions. Table 1 shows autocorrelation lag estimates and pole estimates obtained from HOC and from the traditional batch process, i.e., using (6). The results indicate that the HOC approach provides comparable accuracy to the batch process but with significantly fewer computational operations and less storage.

### B. Application to a closed loop system

Figure 3 shows an example linear discrete-time representation of a closed loop system containing a controller, plant, and noise process. In Figure 2, and in most cases, the control error in a closed-loop system under stochastic regulatory operation can be modeled quite well as an ARMA process.

The closed-loop transfer function between the error signal and the noise input resolves to the following form:

$$\begin{aligned} G_L(z^{-1}) &= \frac{-e_t}{a_t} = \frac{G_n(z^{-1})}{1+G_c(z^{-1})G_p(z^{-1})} \\ &= \frac{ARC}{ARD+DSB} \end{aligned} \quad (21)$$

where $e_t$ is the error signal and $\{a_t\}$ is an IID sequence of Gaussian random variables. Note that the error signal $e_t$ could be replaced by the difference between the controlled variable $y_t$ and its expected value, where the expected value might be estimated.

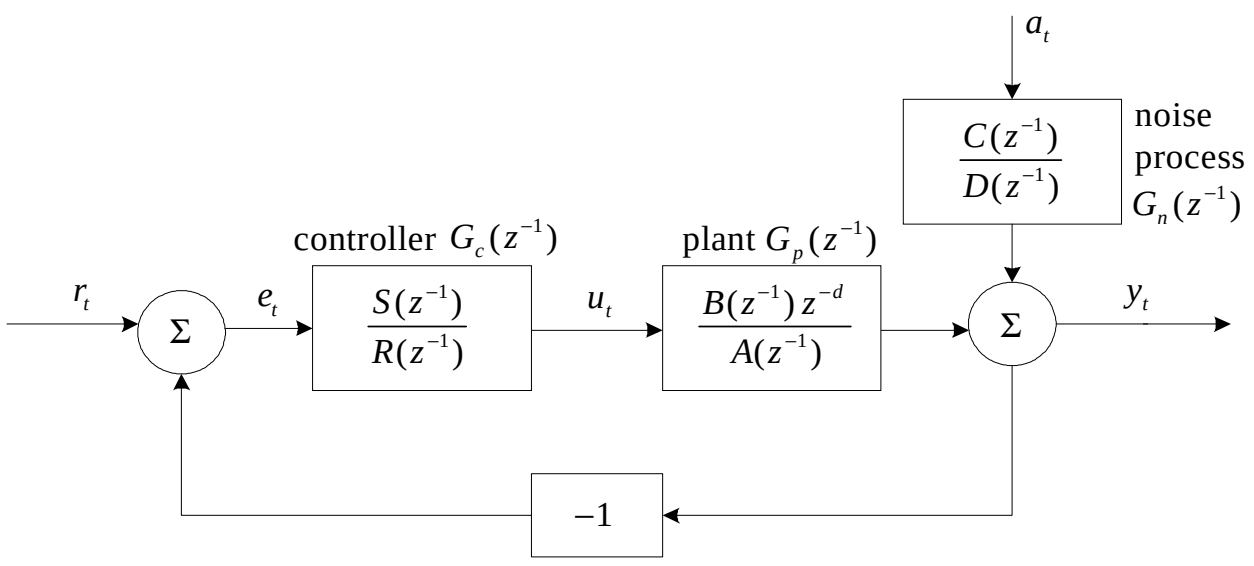

Figure 2: Discrete-time control loop block diagram

For this example, we assume that the noise and plant models are equivalent so that $D \equiv A$ and $C \equiv 1$. For a first-order plus time-delay plant transfer function: $G_p = \dfrac{(1-\alpha)\, z^{-d}}{1-\alpha\, z^{-1}}$, and an integrating controller, $G_c = \dfrac{K_c}{1-z^{-1}}$, the closed loop system becomes an ARMA(MAX(2,d),1) process where:

$$
\begin{aligned}
G_L(z^{-1}) &= \frac{-e_t}{a_t} \\
&= \frac{(1-z^{-1})}{1-(1+\alpha)\,z^{-1}+\alpha\, z^{-2}+(1-\alpha)K_c\, z^{-d}}
\end{aligned}
\tag{22}
$$

Consider now the case where $\alpha = 0.9, d = 2$ in $G_L$ yielding the following second-order system:

$$
G_L(z^{-1}) = \frac{1-z^{-1}}{1-1.9\,z^{-1}+(0.9+0.1K_c)\,z^{-2}} \tag{23}
$$

The method was used to estimate the poles in (23) for different values of $K_c$. Because the process had just two poles, a (continuous-time) damping ratio was calculated to characterize the aggressiveness of the loop. The damping ratio was calculated by assuming that the discrete poles $(p_z)$ map onto their continuous-time counterparts $(p_s)$ through $p_s = \Delta t/\ln(p_z)$, where $\Delta t$ is the sampling period. When the continuous-time poles are complex conjugates such that $p_s = a \pm ib$, the damping ratio is calculated from:

$$
\zeta = \sqrt{\frac{a^2}{a^2+b^2}} \tag{24}
$$

Otherwise, when $\{p_{s,1}, p_{s,2}\}$ are both real,

$$
\zeta = \frac{-(p_{s,1}+p_{s,2})}{2\sqrt{p_{s,1}p_{s,2}}} \tag{25}
$$

The HOC estimation method was tested by generating 10,000 samples from the closed loop system of (23) using four different values for the controller gain: $K_c = [0.05; 0.1; 0.5; 0.75]$. Snapshots of the setpoint error signal are presented in Figure 3 with $K_c$ increasing clockwise from the top left graph.

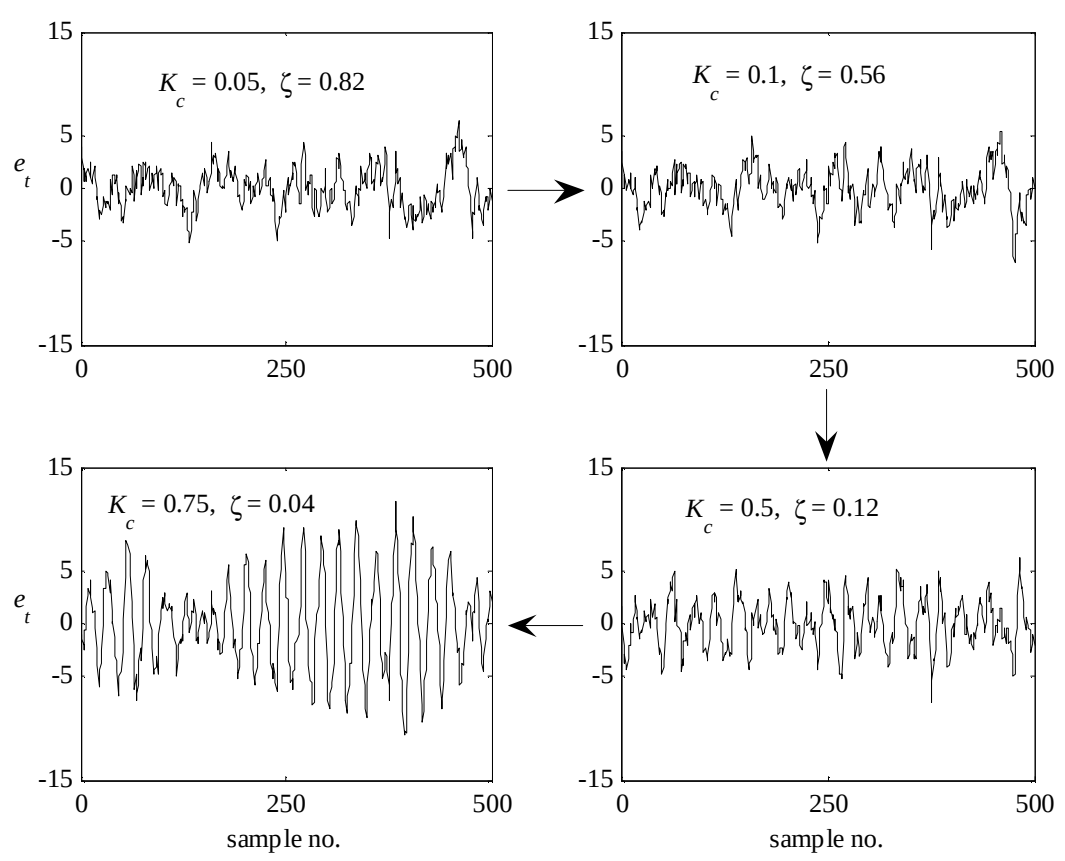

Figure 3: Variation of damping ratio with increasing controller gain

The results of the parameter estimation are shown in Table 2. The actual damping ratio varied from 0.69 to 0.05 and the results show that the accuracy of the parameter estimates improved as the damping ratio became smaller. One reason for this is that peaks in the spectral density function become more pronounced as the damping ratio descreases thereby reducing uncertainty in frequency-based estimation schemes. The results show that the damping ratio, $\zeta$, provides a good indication of aggressiveness and could serve as a way to detect oscillations. For higher-order loops, conjugate pairs of poles could be transformed to damping ratio values to detect the presence of oscillatory modes in the signal. This oscillation detection procedure can be easily implemented online, and requires little process knowledge compared to earlier methods proposed by [26], [27].

Table 2: Results of parameter estimation for closed loop system.

| $K_c$ | Actual Values | | Estimated Values | |
|---|---|---|---|---|
| | $p_{z,\{1,2\}}$ | $\zeta$ | $p_{z,\{1,2\}}$ | $\zeta$ |
| 0.05 | 0.8912±0.0696$i$ | 0.69 | 0.9500±0.0500$i$ | 0.82 |
| 0.1 | 0.9252±0.0980$i$ | 0.46 | 0.9500±0.0866$i$ | 0.56 |
| 0.5 | 0.9482±0.2185$i$ | 0.11 | 0.9500±0.2179$i$ | 0.12 |
| 0.75 | 0.9506±0.2701$i$ | 0.05 | 0.9500±0.2693$i$ | 0.04 |

One issue that needs to be considered when using the proposed method is that of correct order selection. Model order should also be chosen to take into account the possibility of pole-zero cancellation [28].

## VII. CONCLUSION

A new method was presented for estimating the poles in an ARMA process using higher-order crossings. The method involves calculating autocorrelation lags from higher-order crossings counts and solving a set of linear equations to obtain the autoregressive parameters. Poles are obtained by applying a root finding algorithm to the AR parameters. An important advantage of the proposed method is that it does not require storing a batch of data.

The method was tested with three different ARMA processes including an SISO feedback control loop. The results showed that the method provided comparable accuracy to a traditional batch-based approach. The results also demonstrated that the method could be used to detect oscillatory modes in feedback control loops for the purpose of control loop performance assessment.

## REFERENCES

[1] Åström, K. J. 1970. “Introduction to Stochastic Control Theory”. Academic Press, New York and London.

[2] Box, G. E. P., Jenkins, G. M. and Reinsel, G. 1994. “Time-Series Analysis: Forecasting & Control”, 3rd Edition, Prentice Hall, Upper Saddle Rive, NJ.

[3] Hamilton, J. D. 1994. “Time-Series Analysis”. Princeton University Press, Princeton, NJ.

[4] Ljung, L. 1998. “System Identification: Theory for the User”, 2nd Edition. Prentice Hall, Upper Saddle River, NJ.

[5] Brockwell, P. J. and Davis, R. A. 2002. “Introduction to Time-Series and Forecasting”, 2nd Edition. Springer-Verlag, New York.

[6] Harris, T. J. 1989. “Assessment of Control Loop Performance”. Canadian Journal of Chemical Engineering. Volume 67. Page 856.

[7] Kammer, L. C., R. R. Bitmead, P. L. Bartlett. 1998. “Optimal controller properties from closed-loop experiments”. Automatica. Volume 34. Number 1. Page 83-91.

[8] Lee, D. T. L., Friedlander, B. and Morf, M. 1981. “Recursive Ladder Algorithms for ARMA Modeling”. IEEE Trans. Automatic Control, Volume AC-27, Pages 753-764.

[9] Moses, R. L., Cadzow, J. A. and Beex, A. A. 1985. “A Recursive Procedure for ARMA Modeling”. IEEE Trans. Acoust. Speech & Signal Process. Volume ASSP-33, Pages 1188-1196.

[10] Desborough, L., T. Harris. 1992. “ Performance Assessment Measures for Univariate Feedback Control”. The Canadian Journal of Chemical Engineering. Volume 70. December 1992. Pages 1186-1197.

[11] Liao, Y. C. 1989. “Order recursive algorithm for ARMA identification”. In IEEE International Conference on Systems Engineering, Aug 24-26, Fairborn, OH, Pages 209-211.

[12] Efe, M. O., Kaynak, O. and Wilamowski, B. M. 2002. “A Robust Identification Method for Time-Varying ARMA Processes Based on Variable Structure Systems Theory”. Mathematical and Computer Modelling of Dynamical Systems, Volume 8, Pages 185-198.

[13] Strobach, P. 1988. “Recursive Covariance Ladder Algorithms for ARMA System Identification”. IEEE Transactions on Acoustics, Speech, and Signal Processing, Volume 36, Page 560-580.

[14] Friedlander, B., K. Sharman. 1985. “Performance of the Modified Yule-Walker Estimator”. IEEE Transactions on Acoustics Speech and Signal Processing”. Volume ASSP-33. Number 3. Page 719.

[15] Liang, Ying-Chang, Xian-Da Zhang, Yan-Da Li. 1995. “A Hybrid Approach to Time Series Analysis and Spectral Estimation”. Proceedings of the American Controls Conference, Seattle, Washington. Page 124.

[16] Moses, R., P. Stoica, B. Friedlander, T. Söderström. 1987. “An Efficient Linear Method of ARMA Spectral Estimation”. Proceedings of the IEEE. Page 2077.

[17] Pintelon, R., Guillaume, P., Rolain, Y., Schoukens, J. and Van hamme, H. 1994. “Parametric Identification of Transfer Functions in the Frequency Domain”. IEEE Trans. Automatic Control, Volume 39, Pages 2245-2260.

[18] Kac, M. 1943. “On the Average Number of Real Roots of a Random Algebraic Equation”. Bulletin of American Mathematical Society. Volume 49. Pages 314-320.

[19] Rice, S. O. 1945. “Mathematical Analysis of Random Noise”. Bell Systems Technical Journal. Volume 24. Pages 46-156.

[20] Cramér, H., M. R. Leadbetter. 1967. “Stationary and Related Stochastic Processes”. John Wiley & Sons, NY.

[21] Kedem, B. 1980. “Estimation of the Parameters in Stationary Autoregressive Processes After Hard Limiting”. Journal of American Statistical Association. Volume 75. Number 369. Pages 146-153.

[22] Kedem, B. 1986. “Spectral Analysis and Discrimination by Zero-Crossings”. Proceedings of the IEEE. Volume 74. Number 11. Pages 1577-1493.

[23] Kedem, B. 1994. “Time Series Analysis by Higher Order Crossings”. Published by IEEE Press, New York.W.-K. Chen, *Linear Networks and Systems* (Book style). Belmont, CA: Wadsworth, 1993, pp. 123–135.

[24] Barnett, J. T., B. Kedem. 1991. “Zero-Crossing Rates of Functions of Gaussian Processes”. IEEE Transactions on Information Theory. Volume 37. Number 4. Pages 1188-1194.

[25] McFadden, J. A. 1956. “The Axis-Crossing Interval of Random Functions”. IRE Transactions on Information Theory. Volume IT-2. Pages 146-150.

[26] Thornhill, N.F., and Hägglund, T. 1997. “Detection and Diagnosis of Oscillation in control Loops”, Control Engineering Practice, Volume 5, Pages 1343-1354.

[27] Miao, T. and Seborg, D. E. 1999. “Automatic Detection of Excessively Oscillatory Feedback Control Loops”. In Proc. 1999 IEEE Intl. Conf. on Control Applications, Kohala Coas – Island of Hawaii, Aug 22–27, Pages 359-364.

[28] Campi, M. C. 1996. “Problem of Pole-Zero Cancellation in Transfer Function Identification and Application to Adaptive Stabilization”. Automatica. Volume 32. Page 849-857.